\documentclass[aps,pra,twocolumn,superscriptaddress]{revtex4-2}
\usepackage{graphicx}% Include figure files
\usepackage{dcolumn}% Align table columns on decimal point
\usepackage{bm}% bold math
\usepackage{amssymb}
\usepackage{amsmath}
\usepackage{hyperref}% add hypertext capabilities
\usepackage{siunitx}
\usepackage[export]{adjustbox}
\usepackage{upgreek} %upright greek letters
\usepackage{xfrac} %small fractions
\usepackage{physics}
\usepackage{xcolor} %for inkscape colors
\usepackage{accents}
\usepackage{multirow}
\usepackage{booktabs}

\usepackage{subcaption}%replaces subfig
\usepackage{tikz-network} %for coupling graphs
\SetVertexStyle[TextColor = white, TextFont = \tiny, LineWidth=0pt]
\usepackage{tikz}
\usepackage{pgfplots} % LaTeX
\usepgfplotslibrary{colormaps} %colormaps
\usepgfplotslibrary{groupplots} %grouplots
\pgfplotsset{/pgfplots/colormap={PRGn}{rgb255(-1)=(64,0,75) rgb255(0)=(246,246,246) rgb255(1)=(0,68,27)}} %colormap for gradients
\pgfplotsset{/pgfplots/colormap={seismic}{rgb255(-1)=(0,0,255) rgb255(0)=(255,255,255) rgb255(1)=(255,0,0)}} %color map for couplings
\pgfplotsset{/pgfplots/colormap={errorPL}{rgb255(0)=(255,255,255) rgb255(0.46)=(0,255,0)}}
\pgfplotsset{/pgfplots/colormap={errorPLres}{rgb255(0)=(255,255,255) rgb255(0.14)=(0,255,0)}}
\pgfplotsset{/pgfplots/colormap={reds}{rgb255(0)=(255,255,255), rgb255(1)=(255,0,0)}}
\pgfplotsset{/pgfplots/colormap={errorTL}{rgb255(0)=(255,255,255) rgb255(0.33)=(0,255,0)}}
\pgfplotsset{/pgfplots/colormap={errorTLres}{rgb255(0)=(255,255,255) rgb255(0.1)=(0,255,0)}}

\pgfplotsset{compat=1.17}
\pgfkeys{/pgf/number format/.cd,1000 sep={}}



\begin{document}
%% Custom commands

\title{ Engineering spin-spin interactions with optical tweezers in trapped ions}

\author{J. D. Arias Espinoza}\affiliation{Van der Waals-Zeeman Institute, Institute of Physics, University of Amsterdam, 1098 XH Amsterdam, the Netherlands}
\author{M. Mazzanti}\affiliation{Van der Waals-Zeeman Institute, Institute of Physics, University of Amsterdam, 1098 XH Amsterdam, the Netherlands}
\author{K. Fouka}\affiliation{Van der Waals-Zeeman Institute, Institute of Physics, University of Amsterdam, 1098 XH Amsterdam, the Netherlands}
\author{R. X. Sch{\"u}ssler}\affiliation{Van der Waals-Zeeman Institute, Institute of Physics, University of Amsterdam, 1098 XH Amsterdam, the Netherlands}
\author{Z. Wu}\affiliation{Van der Waals-Zeeman Institute, Institute of Physics, University of Amsterdam, 1098 XH Amsterdam, the Netherlands}
\author{P. Corboz}\affiliation{QuSoft, Science Park 123, 1098 XG Amsterdam, the Netherlands}\affiliation{Institute for Theoretical Physics, Institute of Physics, University of Amsterdam, Science Park 904, 1098 XH Amsterdam, the Netherlands}
\author{R. Gerritsma}\affiliation{Van der Waals-Zeeman Institute, Institute of Physics, University of Amsterdam, 1098 XH Amsterdam, the Netherlands}
\author{A. Safavi-Naini}\affiliation{QuSoft, Science Park 123, 1098 XG Amsterdam, the Netherlands}\affiliation{Institute for Theoretical Physics, Institute of Physics, University of Amsterdam, Science Park 904, 1098 XH Amsterdam, the Netherlands}

\begin{abstract}

We propose a new method for generating programmable interactions in one- and two-dimensional trapped-ion quantum simulators. Here we consider the use of optical tweezers to engineer the sound-wave spectrum of trapped ion crystals. We show that this approach allows us to tune the interactions and connectivity of the ion qubits beyond the power-law interactions accessible in current setups. We demonstrate the experimental feasibility of our proposal using realistic tweezer settings and experimentally relevant trap parameters to generate the optimal tweezer patterns to create target spin-spin interaction patterns in both one- and two-dimensional crystals. Our approach will advance quantum simulation in trapped-ion platforms as it allows them to realize a broader family of quantum spin Hamiltonians. 

\end{abstract}

\maketitle

\section{Introduction}

Trapped ions are one of the leading platforms for quantum computation and quantum simulation~\cite{Ballance2016,Gaebler2016,Zhang2017}. Numerous experiments have demonstrated the ability of the analog trapped-ion quantum simulator to emulate the dynamics of quantum magnetism models~\cite{Friedenauer:2008,Lanyon:2011,Cirac:2012,Blatt:2012,Zhang2017} and to study the dynamics of quantum information and quantum entanglement~\cite{Landsman2019, Garttner2017, Lewis-Swan2019}. 

One of the main advantages of the trapped-ion quantum simulators is the tunability of the interaction range, as well as the ability to realize one- and two-dimensional (1D and 2D) systems. Hence, this platform provides an ideal setup in which one can explore the interplay of interaction range and dimensionality in the dynamics of quantum information, entanglement, and speed of thermalization~\cite{Richerme2014, Landsman2019, Garttner2017, Smith2016}, while simultaneously allowing one to simulate models relevant to condensed matter physics that are beyond the state-of-the-art numerical methods. 

However, the current simulator setups do not offer enough versatility to explore the above questions.  This is because, theoretically, the types of engineered interactions which can be realized in these systems is limited to those with power-law decay, $1/r^\xi $, where $r$ is the separation between two ions and $0\leq \xi\leq 3$~\cite{Britton2012}. In order to understand the source of this limitation, we note that the ion-ion interactions in the simulator are phonon-mediated and depend on the spectrum and structure of the collective vibrational modes of the ion crystal~\cite{Porras2004}. Furthermore, considering experimental constraints such as laser power and decoherence rates, the range of interactions tends to be even more limited, and most experiments are operated with $0\leq \xi \leq 1.5$~\cite{Britton2012, Bohnet2016, Richerme2014}.

In the following, we illustrate a new approach for realizing a highly tunable trapped-ion simulator in terms of connectivity, range, and sign of the interactions in both linear (or 1D) and triangular (2D) ion crystals in Paul traps~\cite{Kaufman2012,Landa2012,Wang2014,Nath2015,Richerme2016}. We use optical tweezers to manipulate the frequencies and structure of the collective vibrational modes of the crystal. The triangular crystal structure in 2D makes it a natural platform for implementing quantum simulation of frustrated spin systems~\cite{Bermudez2011,Bermudez2012} and there has been rapid experimental progress in this area in recent years~\cite{Laupretre2019,Qiao2021,Joshi2020}.

\section{Effective Ising Interactions}

The spin-spin interactions in an ion crystal are generated when the electronic state of each ion is coupled to the phonon modes of the ion crystal by applying a state-dependent force generated by a pair of counter-propagating laser beams. The resulting Hamiltonian, in the Lamb-Dicke (LD) limit and in the interaction picture with respect to $\hat H_0=\sum_{m=1}^N \hbar \omega_m \hat n_m$, is given by
\begin{align}
\label{eq:HI}
    \nonumber\hat{H}_{\rm I} = -i\sum_{j=1}^N &\,g \cos(\mu t)\hat{\sigma}_j^x \\& \times \sum_{m=1}^N b_{jm} \sqrt{\frac{\hbar}{ 2M \omega_m}} \left(\hat a_m e^{-i \omega_m t} - \hat a_m^\dagger e^{i \omega_m t} \right). 
\end{align}
where where $\mu$ is the beat-note frequency between the two counter-propagating bichromatic lasers used to generate the state-dependent force and $g$ is the interaction strength of the laser which is assumed to be homogeneous throughout the crystal. Here $\hat a_m$ ($\hat a_m^\dagger$) is the annihilation (creation) operator for a phonon in mode $m$, $\hat n_m=\hat a_m^\dagger \hat a_m$, and $b_{jm}$ is the amplitude of the corresponding eigenvector at the position of ion $j$. The LD parameter is given by $\eta_j^{(m)}\equiv k_q b_{jm}\sqrt{\frac{\hbar}{2M \omega_m}}$ where $k_q$ is the wavevector for the Raman beam pair and $M$ the ion mass. 

The form of the Hamiltonian in Eq.~\eqref{eq:HI} allows us to find an explicit expression for the propagator governing the evolution of the system. The propagator can be written explicitly as its Magnus series expansion truncates at second order~\cite{Porras2004, Sorensen1999, Monroe2009}: 
\begin{align}
    \hat{U}(t,0) &= \exp\Bigg(-\frac{i}{\hbar}\int_0^t dt^\prime \hat{H}_I(t^\prime) \nonumber\\ 
    &-\frac{i}{2\hbar}\int_0^t dt^\prime \int_0^{t^\prime} dt^{\prime\prime} \left[\hat{H}_I(t^\prime),\hat{H}_I(t^{\prime\prime})\right]\Bigg)\\
    &\approx \exp\Bigg(i\sum_j\left[\gamma_j^{(m)}(t)\, \hat{a}_m +h.c. \right]\hat{\sigma}_j^x \nonumber\\
    &-i\sum_{j,k} \beta_{j,k}(t)\hat\sigma_j^x \hat{\sigma}_k^x \Bigg), 
\end{align}
where, 
\begin{equation*}
\gamma_j^{(m)}(t)=\frac{-i g \eta_{j}^{(m)} }{\mu^2-\omega_m^2}\\
\bigg[\mu
 -e^{i \omega_m t}\left(\mu \cos\left(\mu t\right)-i\omega_m \sin\left(\mu t\right)\right)\bigg]
\end{equation*}
and 
\begin{align*}
\beta_{j,k}(t)=-g^2\sum_m& \frac{ \eta_{j}^{(m)} \eta_{k}^{(m)}}{ \mu^2-\omega_m^2} \bigg(\frac{\mu \sin((\mu-\omega_m)t)}{\mu-\omega_m} \\
- &\frac{\mu \sin((\mu+\omega_m)t)}{\mu+\omega_m} +\frac{\omega_m\sin(2\mu t)}{2\mu}-\omega_m t\bigg).
\end{align*}

We note that the term containing the spin-spin interaction in the propagator $\hat U(t,0)$ grows linearly with time and can be identified as the phase of the time evolution of a system evolving with an effective Ising Hamiltonian \(\hat{H}_{\rm Ising} = \sum J_{j,k}\hat{\sigma}_j^x\hat{\sigma}_k^x \) with 
\begin{equation}
\label{eq:Jij}
J_{j,k}=  g^2\sum \omega_m \frac{\eta_{j}^{(m)} \eta_{k}^{(m)} }{\mu^2-\omega_m^2}. 
\end{equation}

As it is shown in Eq.~\eqref{eq:Jij}, the interactions between the ions are determined by the structure of the phonon modes. Thus, the ability to engineer these collective vibrational modes of the crystal allows us to realize a wide variety of interaction matrices.

Our method uses local optical potentials to induce additional (anti-)confinement of individual ions and frequency state-dependent Raman forces. The result of the additional optical potentials is a change in the phonon mode spectra and of the individual amplitudes of each mode at each ion. 

The main result of our work is as follows: given a target 1D or 2D spin-\(\tfrac{1}{2}\) Hamiltonian, specified by the interaction matrix $\mathbf J$,  of the form

\begin{equation}\label{eq:general_hamiltonian}
    \hat{H}_{\rm T} = \sum_\alpha \sum_{j<k} J_{j,k}^{\alpha} \hat\sigma_i^{\alpha} \hat\sigma_j^{\alpha}, % + \sum_{i,\alpha} h^{(i)}_\alpha \hat\sigma^{(i)}_\alpha
\end{equation}
we (a) outline a procedure to determine if the target Hamiltonian can be realized, and (b) provide a systematic approach to find the optimal tweezer pattern to realize $\hat H_{\rm T}$. Here we have used $J_{j,k}^\alpha$ to denote the matrix elements of the target interaction matrix $\mathbf J$  along the Cartesian coordinate $\alpha=x, y, z$.

\section{Phonon mode engineering}

In order to characterize the effect of the tweezer potential on the normal modes, we consider $N$ ions of mass $M$ which are confined by a harmonic trapping potential $V_{\rm trap} \left(\bm \rho_i\right)=\frac{1}{2}\sum_{\alpha,i} M\omega_{\alpha}^2 \rho_{i,\alpha} ^2$, where $\bm \rho_i=\left(\rho_{i,x}, \rho_{i,y}, \rho_{i,z} \right)$ is the position of the $i$-th ion and $\omega_\alpha$ is the trap frequency in the $\alpha$ direction. The ions are further confined by a tweezer potential $V_{\rm tweezer}$. Thus, the potential energy of the system is given by: 
\begin{equation}
\label{eq:Vpot1}
    V(\bm \rho_i)=V_{\rm trap}(\bm \rho_i)+V_{\rm tweezer}(\bm \rho_i)+\frac{1}{2}\sum_{i\neq j}\frac{e^2}{4\pi\epsilon_0\vert \bm \rho_i-\bm \rho_j\vert}, 
\end{equation}
\noindent
where the third term is the Coulomb potential between the ions, $e$ the Coulomb constant, and $\epsilon_0$ is the vacuum permittivity.  

In order to find the collective vibrational modes of the ion crystal we follow the procedure described in~\cite{James:1998} and find the equilibrium positions $\mathbf r_i^{(0)}$ corresponding to the solutions of $\nabla V=0$.
Assuming that the ions perform small oscillations about this equilibrium position, the position of each ion can be written as $\bm \rho_i=\mathbf r_i^{(0)}+ \mathbf r_i$, where $\mathbf r_i$ are deviations from the equilibrium position. Furthermore, considering the tweezer intensity profile at each ion, $I(\mathbf r_i)$, being centered at each corresponding equilibrium position $\mathbf r_i^{(0)}$, the optical dipole potential can be approximated as harmonic. In this way, the general form of the optical tweezers potential is given as:

\begin{equation}\label{tweezers}
    V_{\rm tweezer}(\mathbf r_i) = \sum_{i=1}^N \sum_{\alpha, \alpha'}\frac{M}{2}  \Omega_{i,\alpha,\alpha'}^2\left(\alpha_i - \alpha_i^{(0)} \right)\left(\alpha'_i - {\alpha'}_i^{(0)} \right),
\end{equation}
where $\Omega_{i,\alpha,\alpha'}^2$ denotes the local optical pinning curvature expressed as the square of the trap frequency squared at the $i$-th ion. $\alpha_i$ and $\alpha_i^{(0)}$ denote the ion position and equilibrium position along the corresponding direction.

The addition of $V_{\rm tweezer}$ does not modify the equilibrium position of the ions since it vanishes there. However, the collective modes of the crystal are modified by the additional tweezer potentials. We obtain the phonon spectrum by expanding the Coulomb potential to second order in $\mathbf r_i$. The resulting Lagrangian, using the explicit form of $V_{\rm tweezer}$ given in Eq.~\eqref{tweezers}, is: 

\begin{align}
    \mathcal{L}=&\frac{M}{2}\Bigg(\sum_i\sum_\alpha\left(\dot \alpha_i\right)^2 \nonumber\\ 
    -&\frac{1}{2}\sum_{i,j}\sum_{\alpha,\alpha'} \alpha_i \alpha'_j\left(\frac{d^2V}{d\alpha_i d\alpha'_j}\right)_{\alpha_i, \alpha'_j\rightarrow 0}\Bigg) \nonumber \\
    =&\frac{M}{2}\left(\sum_i\sum_\alpha\left(\dot\alpha_i\right)^2-\frac{1}{2}\sum_{i,j}\sum_{\alpha,\alpha'} \alpha_i \alpha'_j A^{(i,j)}_{\alpha,\alpha'}\right).
\end{align}
The eigenvalues and eigenvectors of the Hessian matrix $\mathbf A$ determine the normal modes of the crystal. The $k$-th normalized eigenvector corresponds to a normal mode of the crystal, which we denote as $\mathbf b_k$. Its entries represent the amplitude of motion of each ion in each of these modes. The modes frequencies are $\omega_k=\sqrt{\lambda_k}$ with $\lambda_k$ the eigenvalues of $\mathbf A$. In case of a 1D ion crystal, the eigenmodes separate in three subclasses, corresponding to the directions of motion $x,y,z$ and $k=1, \dots, 3N$.

\section{Optimization of non-native spin-spin interactions}\label{sec:opt}

The realization of the target Hamiltonian given by $\hat{H}_\text{T}$ of the form Eq.~\eqref{eq:general_hamiltonian} with specified coupling matrix $\mathbf J_T$, relies on finding the optimal confinement realized by the tweezer at the position of each ion. We solve this problem through three optimization steps. 

First, we consider an equidistantly spaced ion crystal with inter-ion distance \(d_0\), defined by an effective axial trap frequency $\omega_{z,\text{eff}}$ as~\cite{James:1998,Lin2009}:

\begin{equation*}
    d_0 \approxeq \left(\frac{e^2}{4\pi\epsilon_0 M \omega_{z,\text{eff}}^2}\right)^{\frac{1}{3}}\frac{2}{N^{0.56}}.
\end{equation*}

We search for values of \(\mathbf{s}^\prime=\{\omega_z,\;\mu^\prime,\;\Omega^\prime_i\}\) that minimize the error of the matrix of Ising couplings:

\begin{equation}
\label{eq:Jerr}
    \epsilon(\mathbf{s}) = \frac{\left\Vert \mathbf{J}_\text{T} - \mathbf{\tilde J}(\mathbf{s}) \right\Vert}{\lVert \mathbf{J}_\text{T}\rVert}.
\end{equation}

Here \(\mathbf{\tilde J}(\mathbf{s})\) is the resulting coupling matrix of the pinned crystal, normalized such that the largest entries of \(\mathbf{J}_\text{T}\) and \(\mathbf{\tilde J}\) have the same magnitude. This problem is formulated as constrained optimization of the form:
\begin{equation}\label{eq:optimization_step1}
    \underset{\mathbf{s}}{\text{argmin}}(\epsilon(\mathbf{s}))\;: \;s_i^\text{min}\leq s_i \leq s_i^\text{max}.\\
\end{equation}

\begin{figure}
    \begin{tikzpicture}
        \node[matrix,draw=red] (J0) at (-2,0)
        {
            & \node{+}; & \node{--}; & \node{+}; & \node{+}; \\
            \node{+}; &  & \node{+}; & \node{--}; & \node{--}; \\
            \node{--}; & \node{+}; &  & \node{--}; & \node{--}; \\
            \node{+}; & \node{--}; & \node{--}; &  & \node{--}; \\
            \node{+}; & \node{--}; & \node{--}; & \node{--}; &  \\
        };
        \node[matrix,draw=blue] (Jt) at (2,0)
        {
             & \node{+}; & \node{--}; & \node{--}; & \node{--}; \\
            \node{+}; &  & \node{+}; & \node{--}; & \node{--}; \\
            \node{--}; & \node{+}; &  & \node{+}; & \node{--}; \\
            \node{--}; & \node{--}; & \node{+}; &  & \node{+}; \\
            \node{--}; & \node{--}; & \node{--}; & \node{+}; & \\
        };
        \draw [->,black,thick] (J0.east) -- (Jt.west) node [above,midway] {$\exists \;\mathbf{\Omega}$ ?};
    \end{tikzpicture}
    \caption{\label{fig:sign_structure} Determining the feasibility of a target coupling graph. For a fixed crystal geometry and beatnote frequency, we determine if a set of optical potential frequencies exists that fulfill the gradient condition.}
\end{figure}
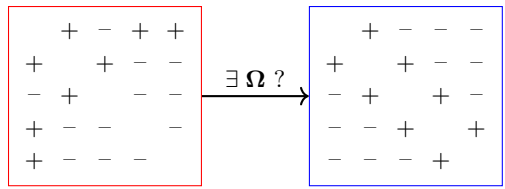
%we note that our method works best when the couplings matches to a symmetry of the crystal, i.e. when the distance between the eigenmodes of \(J^{(ij)}_\alpha\) and the eigenmodes (phonon modes) of the crystal is small.
We limit the parameter space of the search by first performing a feasibility test (Fig. \ref{fig:sign_structure}) for the sign structure of \(J^{(ij)}_\alpha\) for values of \(\{\omega_z,\;\mu\}\). This is motivated by noting that when the target couplings match the sign structure of the couplings of the unpinned crystal, the optimization is more successful. 
The test consists in determining the existence of a solution \(\mathbf{\Omega}\) which satisfies the inequality \(\mathbf{X}\mathbf{\Omega} > 0\) holds, where \(\mathbf{X}=\left(\mathbf X^{(1)}, \mathbf X^{(2)}, \cdots \right)^T\) is a matrix whose rows are gradient constraints for each coupling strength of interest in \(\mathbf{J}_\text{T}\). Each row \(\mathbf{X^{(c)}}\) is expressed as:

\begin{equation}\label{eq:gradient_condition}
    \mathbf{X^{(c)}}=\text{sign}(\Delta \text{J}_{k,l})\left(\bar{A}^{(1,1)}, \cdots,\; \bar{A}^{(N,N)}\right),
\end{equation}

\noindent where \(\mathbf{\Delta J} = \mathbf{J}_\text{T} - \mathbf{J}_0\), \(\mathbf{J}_0\) is the coupling matrix of the unmodified system and \(\bar{A}^{(i,i)}\) is the gradient of \(\mathbf{J_0}\) with respect the diagonal matrix element of the Hessian \(A^{(i,i)}\) (see Appendix). If a solution exists for a particular pair \(\{\omega_z,\;\mu\}\), we perform a second optimization only for the optical potential frequencies:

\begin{equation}\label{eq:optimization_step2}
    \underset{{\Omega^\prime_i}}{\text{argmin}}(\epsilon(\mathbf{s}))\;: \; \Omega_i^\text{min}\leq \Omega^\prime_i \leq \Omega_i^\text{max}.\\
\end{equation}

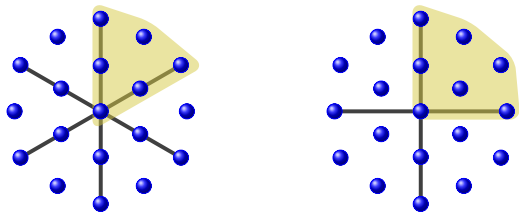
\begin{figure}
    \begin{subfigure}{0.23\textwidth}
        \begin{tikzpicture}[scale=0.5]
            \fill[fill=yellow!80!black, fill opacity=0.5, rounded corners] (-0.22, -0.5) -- (-0.22, 2.9) -- (1.3, 2.4) -- (2.75, 1.2) -- cycle;
            \Vertices[NoLabel, size=.2, opacity=.9, style={shading=ball,blue}]{Data/AFM_triangularvertices.csv}
            \Edge(1)(19)
            \Edge(1)(17)
            \Edge(1)(15)
            \Edge(1)(13)
            \Edge(1)(11)
            \Edge(1)(9)
        \end{tikzpicture}
    \end{subfigure}
        \begin{subfigure}{0.23\textwidth}
        \begin{tikzpicture}[scale=0.5]
            \fill[fill=yellow!80!black, fill opacity=0.5, rounded corners] (-0.22, -0.22) -- (-0.22, 2.9) -- (1.3, 2.4) -- (2.5, 1.4) -- (2.65, -0.22) -- cycle;
            \Vertices[NoLabel, size=.2, opacity=.9, style={shading=ball,blue}]{Data/AFM_triangularvertices.csv}
            \Edge(1)(19)
            \Edge(1)(16)
            \Edge(1)(10)
            \Edge(1)(13)
        \end{tikzpicture}
        \end{subfigure}
    \caption{\label{fig:symmetries_optimization} Examples of unit cell for two hypothetical coupling graphs. Only the values of the optical potentials of the ions in the shaded areas have to be optimized. The potentials at the remaining locations are related by rotations to the potentials of the unit cell.}
\end{figure}

This optimization step can be simplified by considering the rotational symmetries of \(\mathbf{J}_\text{T}\) in the crystalline lattice. We thus define a rotational unit cell for the lattice and optimize only for optical potentials of that cell (Fig. \ref{fig:symmetries_optimization}).

In the final step, we calculate the actual positions of the ions in a harmonic trap for the optimal value \(\omega_z^\prime\). Next, we perform a second search of optimal values of \(\mathbf s = \{\mu, \Omega_i\}\) using as initial guess the values of \({\mu^\prime, \Omega^\prime_i}\) found in the previous steps.

\section{Examples}

For the results presented in this section, we have used a pseudo-hessian method (L-BFGS) in combination with a line-search scheme (Backtracking or Hager Zhang) to perform the optimizations of Eq. \ref{eq:optimization_step1} and Eq. \ref{eq:optimization_step2}. In general, convergence is obtained in much less than one minute for the Backtracking scheme or a few minutes for the Hager Zhang scheme using a desktop computer \footnote{We have used packages Optim.jl and JuMP.jl version 1.2.0 and 0.21.6 respectively loaded in Julia version 1.5.2}.

\subsection{1D crystals}

We first consider a linear chain of ions. A 1D crystal of $N$ ions features $3N$ phonon modes: $N$ axial phonon modes and 2$N$ radial modes. The radial modes are split into two groups of $N$ with an identical set of $N$ eigenvectors $\lbrace \mathbf b^m\rbrace_{m=1}^N$. We note that it is possible to generate Ising-like interactions by coupling to modes belonging to each of the three sets. Hence, engineering all $3N$ phonon modes may allow for a more flexible quantum simulator. However, for simplicity, we first consider only coupling to one group of radial modes, e.g. those along the $y$-direction.

% Homogenous coupling figure
\begin{figure}
    \begin{subfigure}{0.23\textwidth}
        \subcaptionbox{\label{label1}}
        {
            \begin{tikzpicture}[baseline]
            \begin{axis}[
            unit vector ratio=1 1 8,
            every axis z label/.style={at={(ticklabel cs:1.25)}, rotate=0, anchor=west,},
            view = {110}{30},% important to draw x,y in increasing order
            small,
            zlabel = $\tilde{J}_{i,j}$,
            xmin = 0,
            ymin = 0,
            xmax = 12,
            ymax = 12,
            zmin = 0,
            xticklabels= {0,1,6,12},
            yticklabels= {0,1,6,12},
            unbounded coords = jump,
            colormap={pos}{color(0cm)=(white); rgb255(1cm)=(127,0,0)} %see pg.127 pgf manual
            ]
            \addplot3[surf,
            mark=none,
            fill opacity=1.0,
            shader=flat corner,
            draw=black,
            draw opacity=0.5,
            line width=0.05pt
            ] coordinates {\input{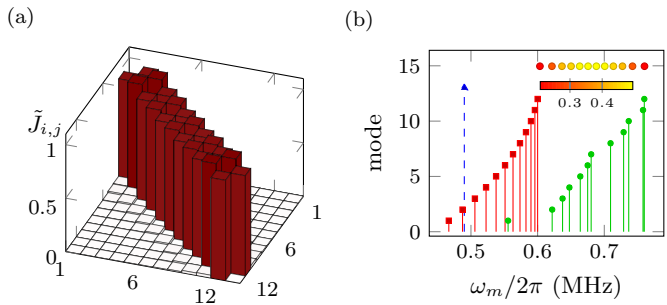} };
            \end{axis}
            \end{tikzpicture}
        }
    \end{subfigure}
    \hspace{0.15cm}
    \begin{subfigure}{0.23\textwidth}
        \subcaptionbox{\label{label2}}
        {
            \begin{tikzpicture}[baseline]
                \begin{axis}[
                    xlabel = $\omega_m/2\pi$ (MHz), 
                    ylabel = mode,
                    small, 
                    width=4.7cm
                    ]
                    \addplot+ [ycomb, mark size = 1.5pt, mark=triangle*, dashed] coordinates {(0.49,13)};%beatnote \mu/2\pi
                    \addplot+ [ycomb, mark size = 1pt] table [x index=1, y index=0, col sep=comma] {Data/Homogeneous_coupling_mode_spectra.dat};
                    \addplot+ [ycomb, mark size = 1pt, mark options={green!80!black}, color=green!80!black] table [x index=2, y index=0, col sep=comma] {Data/Homogeneous_coupling_mode_spectra.dat};
                    \addplot+ [mark size = 0pt, mark=triangle*] coordinates {(0.6,15)};
                \end{axis}
                \begin{axis}[
                    hide axis,
                    tiny,
                    width=3.25cm,
                    xshift = 38pt, 
                    yshift = 47pt,
                    colorbar horizontal,
                    colormap/redyellow,
                    colorbar style={height=3pt, width=35pt, yshift=4pt,xshift=4pt}
                    ]
                    \addplot+ [scatter, scatter src=explicit, only marks, mark options={draw=none}, mark size = 1.25pt, colormap/redyellow] table [x index=1, y index=2, meta index=3, col sep=comma] {Data/Homogeneous_coupling_tweezer_strength.dat};
                \end{axis}
            \end{tikzpicture}
        }
    \end{subfigure}
    \caption{\label{fig:homogeneous_coupling} Nearest-neighbor homogeneous interaction for a linear ion crystal. (a) Resulting interaction matrix and (b) mode spectra of modified ($\textcolor{green}{\bullet}$)  and native ($\scriptscriptstyle\textcolor{red}{\blacksquare}$) phonon modes. The Raman beatnote is indicated by ($\scriptstyle\textcolor{blue}{\blacktriangle}$). Inset: The optical tweezer strength ($\Omega_x^{(i)}/2\pi$ (MHz)) at each ion.}
\end{figure}

\begin{figure}
    \begin{subfigure}{0.23\textwidth}
        \subcaptionbox{\label{label1}}
        {
            \begin{tikzpicture}[baseline]
            \begin{axis}[
            unit vector ratio=1 1 2,
            every axis z label/.style={at={(ticklabel cs:1.25)}, rotate=0, anchor=center,},
            view = {110}{30},% important to draw x,y in increasing order
            small,
            zlabel = $\log \tilde{J}_{i,j}$,
            xticklabels= {0,1,6,12},
            yticklabels= {0,1,6,12},
            xmin = 0,
            ymin = 0,
            xmax = 12,
            ymax = 12,
            zmin = 0,
            unbounded coords = jump,
            colormap={pos}{color(0cm)=(white); rgb255(1cm)=(127,0,0)} %see pg.127 pgf manual
            ]
                \addplot3[
                    surf,
                    mark=none,
                    fill opacity=1.0,
                    shader=flat corner,
                    draw=black,
                    draw opacity=0.5,
                    line width=0.05pt
                    ] coordinates {\input{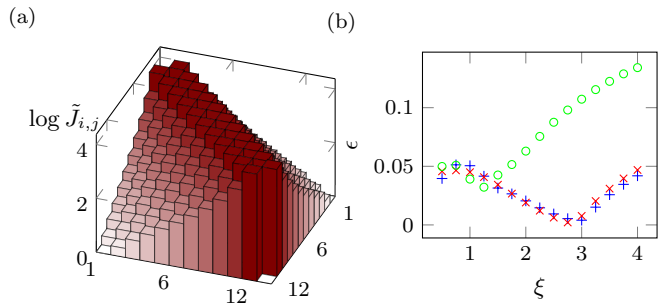} };
            \end{axis}
            \end{tikzpicture}
        }
    \end{subfigure}
        %\hspace{0.15cm}
    \begin{subfigure}{0.23\textwidth}
        \subcaptionbox{\label{label2}}
        {
        \begin{tikzpicture}[baseline]
            \begin{axis}[
                %scaled y ticks = false,
                %every axis y label/.style={at={(ticklabel cs:0.5)}, rotate=90, anchor=center,},
                small,
                width=4.7cm,
                xlabel = $\xi$,
                ylabel = $\epsilon$,
                yticklabels = {0,0,0.05,0.1},
                legend style={draw=none}
                ]
            \addplot+[only marks, mark = +, mark size = 2pt] table [x index=0, y index=1]{Data/error_power_law_even.txt};
            \addplot+[only marks, mark = x, mark size = 2pt] table [x index=0, y index=1]{Data/error_power_law_uneven.txt};
            \addplot+[only marks, mark size = 1.5pt, mark = o, draw =green] table [x index=0, y index=1]{Data/error_power_law_uneven_nopin.txt};
            \end{axis}
        \end{tikzpicture}
        }
    \end{subfigure}
    \caption{\label{fig:power_law} (a) An example of a power-law interaction \(\frac{1}{r}\) coupling matrix for a linear crystal with unequal spacing. (b)~Error \(\epsilon=\lVert \mathbf{J}_\text{T} - \mathbf{\tilde{J}} \rVert/\lVert \mathbf{J}_\text{T} \rVert\) of the resulting coupling matrices for different power-law strengths for equally (\textcolor{blue}{+}) , unequally (\textcolor{red}{$\times$}) spaced linear crystals, and without tweezers (\textcolor{green}{$\circ$}).}
\end{figure}

We demonstrate the tunability of interactions using phonon mode engineering by considering a linear crystal of 12 ions in a Paul trap and two target interaction forms: (i)~a homogeneous nearest-neighbor interaction (Fig.~\ref{fig:homogeneous_coupling}), and (ii) a controllable power-law interaction (Fig. \ref{fig:power_law}(a)), both with anti-ferromagnetic (AF) couplings. We summarize the relevant experimental parameters for the optimal configuration for both scenarios in 1D in Table~\ref{tab:summary_1D} and discuss our choices further in section~\ref{sec:exppar}.

A nearest-neighbour interaction is the anti-thesis of the phonon-mediated interactions arising in the trapped-ion quantum simulators in the absence of the local pinning potentials introduced in this work. The collective nature of the phonon modes give rise to effective spin-spin interactions that are long-range in character. However, as we show in Fig.~\ref{fig:homogeneous_coupling}(a), using experimentally accessible parameters, we can generate a uniform, nearest-neighbour AF coupling matrix. In  Fig.~\ref{fig:homogeneous_coupling}(b) we show how the addition of the pinning potential, shown as an inset, shifts the original normal modes frequencies. 

In the case of the power-law couplings, we consider an ion crystal in a segmented Paul trap with almost equidistant ion spacing and compare it against a crystal in a harmonic trap where the spacing between ions decreases away from the center. Finally we compare both scenarios to a crystal in a harmonic trap in the absence of the tweezer potential. For this scenario we use the same trap parameters but vary the beatnote frequency $\mu$ to minimize the error as defined in section~\ref{sec:opt}.

Figure~\ref{fig:power_law}(b) shows that, in the presence of the tweezer potential, for both evenly spaced crystals and crystals in harmonic traps, the smallest error of the approximated power-law decay occurs close or at \(\xi=3\) which corresponds to interactions of dipole-dipole nature~\cite{Porras2004}. Furthermore, we find that for shorter-range interactions with $\xi \sim 1.5$ the addition of tweezers significantly reduces the error between the target Hamiltonian and the realized Hamiltonian.

% Power law figure

\begin{table}
\caption{\label{tab:summary_1D} Summary of frequencies for the homogeneous near-neighbour (NN) and two values
of power-law couplings in equally (\textcolor{blue}{+}) and unequally (\textcolor{red}{$\times$}) spaced (harmonically trapped) 1D crystals. The highlighted values indicate the direction of the confinement.}
\small
\centering
\begin{tabular}{cl>{\centering\arraybackslash}p{2.1cm}c>{\centering\arraybackslash}p{2.3cm}c>{\centering\arraybackslash}p{2.3cm}c}

 \textbf{Freq.}  & & \multirow{2}{*}{\textbf{NN}} & \multirow{2}{*}{\(\mathbf{\xi}=3.5\)}  &  \multirow{2}{*}{\(\mathbf{\xi}=1.5\)}\\ 
 (MHz$/2\pi$)\\
\toprule

\multirow{2}{*}{\(\mathbf{\omega}_\alpha\)} & \textcolor{blue}{+} & - & \textbf{0.6}, 0.6, 0.33 & \textbf{0.6}, 0.6, 0.33\\ %even 
 & \textcolor{red}{$\times$} & 2, \textbf{0.6}, 0.07 & \textbf{0.6}, 0.6, 0.1 & \textbf{0.6}, 0.6, 0.1 \\ %uneven
 \midrule
 \multirow{2}{*}{max \(\Omega_\alpha^{(i)}\)} & \textcolor{blue}{+}& - & 2.0 & 0.4 \\ %even
 & \textcolor{red}{$\times$} & 0.5  & 2.0 &  0.95 \\ %uneven
 \midrule
 \multirow{2}{*}{\(\mu\)} & \textcolor{blue}{+} &- & 4.6  &  0.8 \\ %even

& \textcolor{red}{$\times$} & 0.49  & 2.9  &  1.2 \\ %uneven

\bottomrule

\end{tabular}
\end{table}

\subsection{2D crystals}

\begin{table}
\caption{\label{tab:summary_2D} Summary of frequencies for the spin-ladder (SL) and nearest-neighbour triangular lattice (TL). Highlighted is the direction of the optical potentials}
\small
\centering
\begin{tabular}{>{\centering\arraybackslash}p{2.1cm}c>{\centering\arraybackslash}p{2.6cm}c>{\centering\arraybackslash}p{2.6cm}c}

 \textbf{Freq.} & \multirow{2}{*}{ \textbf{SL}} &\multirow{2}{*}{ \textbf{TL}} \\
 (MHz$/2\pi$) \\
\toprule
 \(\mathbf{\omega}_\alpha\) &  0.6, \textbf{0.4}, \textbf{0.14} & \textbf{2.4}, 0.16, 0.16\\ 
\midrule
max \(\Omega_\alpha^{(i)}\) &  0.7 & 0.29\\
\midrule
\(\mu\) &  4.2  &  2.4 \\
\bottomrule
\end{tabular}
\end{table}

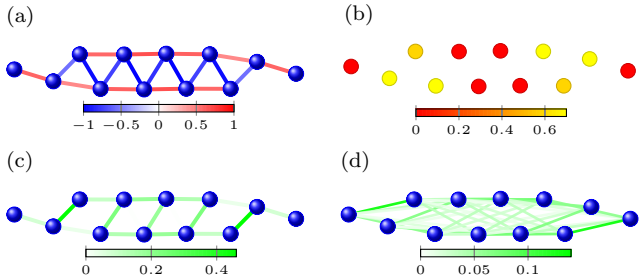
\begin{figure}
    \begin{subfigure}{0.23\textwidth}
        \subcaptionbox{\label{label1}}
        {\begin{tikzpicture}[baseline]
            \SetDistanceScale{0.65}
            \Vertices[size=.2, opacity=.7, style={shading=ball,blue}]{Data/Spin_Ladder_LBFGSvertices.csv}    
            \Edges[RGB]{Data/edges_spin_ladder_12.csv}    
        \end{tikzpicture}}
        \pgfplotscolorbardrawstandalone[colormap={}{of colormap={seismic,},},
        point meta min=-1,
        point meta max=1,
        colorbar horizontal,
        colormap access=map,
        colorbar style={
        width=2cm, height=0.1cm, font=\tiny, yshift=-20pt},
        ]
    \end{subfigure}
    ~
    \begin{subfigure}{0.23\textwidth}
        \subcaptionbox{\label{label2}}
        {\begin{tikzpicture}[baseline]
            \begin{axis}[
                hide axis,
                colorbar horizontal,
                colormap/redyellow,
                width=6cm,
                height=2.15cm,
                colorbar style={width=2cm, height=0.1cm, font=\tiny, anchor=south west, xshift=35pt},
                ]
                \addplot+ [scatter, scatter src=explicit, only marks, mark options={draw=none}, mark size = 2.75pt, colormap/redyellow] table [x index=1, y index=2, meta index=3, col sep=comma] {Data/Spin_Ladder_LBFGS_tweezer_strength.dat};
            \end{axis}
        \end{tikzpicture}}
    \end{subfigure}
    ~
    \begin{subfigure}{0.23\textwidth}
        \subcaptionbox{\label{label3}}
        {\begin{tikzpicture}[baseline]
            \SetDistanceScale{0.65}
            \Vertices[size=.2, opacity=.7, style={shading=ball,blue}]{Data/Spin_Ladder_LBFGSvertices.csv}
            \Edges[RGB,NoLabel,lw=1.5, opacity=1]{Data/Spin_Ladder_LBFGS_abs_erroredges.csv}  
        \end{tikzpicture}}
        \pgfplotscolorbardrawstandalone[colormap={}{of colormap={errorPL,},},
        point meta min=0,
        point meta max=0.46,
        colorbar horizontal,
        colormap access=map,
        colorbar style={
        width=2cm, height=0.1cm, font=\tiny},
        ]
    \end{subfigure}
    ~
    \begin{subfigure}{0.23\textwidth}
        \subcaptionbox{\label{label3}}
        {\begin{tikzpicture}[baseline]
            \SetDistanceScale{0.65}
            \Vertices[size=.2, opacity=.7, style={shading=ball,blue}]{Data/Spin_Ladder_LBFGSvertices_residual.csv}
            \Edges[RGB,lw=1.0, opacity=0.9]{Data/Spin_Ladder_LBFGS_erroredges_residual.csv}  
        \end{tikzpicture}}
        \pgfplotscolorbardrawstandalone[colormap={}{of colormap={errorPLres,},},
        point meta min=0,
        point meta max=0.14,
        colorbar horizontal,
        colormap access=map,
        colorbar style={
        width=2cm, height=0.1cm, font=\tiny, /pgf/number format/fixed},
        ]
    \end{subfigure}
    \caption{\label{fig:ladder} a) Nearest-neighbor couplings in a spin-ladder with frustration, b) optical tweezer strength ($\Omega_{y,z}^{(i)}/2\pi$ (MHz)) at each ion and deviation in coupling strength (\(\lvert J_\text{T,i,j} -\tilde{J}_{i,j} \rvert\))
    for (c) nearest neighbor couplings and (d) residual couplings}
\end{figure}

We now consider 2D crystals, with $N$ transverse phonon modes and $2N$ in-plane modes. While it is easier to isolate single transverse modes to generate the desired interaction, the planar modes offer more degrees of freedom and thus allow us to engineer a wider variety of couplings. We illustrate this using the spin-ladder with frustration which is shown in Fig.~\ref{fig:ladder}(a), for a 2D crystal in a harmonic trap. Here nearest neighbouring particles interact ferromagnetically along the rungs of the ladder and anti-ferromagnetically along the legs of ladder. This type of interaction is best engineered through the in-plane modes since one such mode has the underlying sign structure to realize this interaction graph. The addition of optical tweezers in the optimal configuration corrects the amplitude of the dominant phonon modes at the ion positions such that the desired couplings are achieved. Fig.~\ref{fig:ladder}(c) and Fig.~\ref{fig:ladder}(d) characterize the two sources of errors in the realized Ising couplings: (i) the non-uniformity of the nearest-neighbour couplings, and (ii) residual longer-ranged couplings. 

We note that our optimization uses a particular error function. However, depending on the application of the simulator other error functions may be more suitable. For instance, some physical phenomena may be robust to the presence of longer-range interactions but sensitive to the non-uniform couplings. Thus a case-specific function can be used that favours one particular outcome of the optimization according to the needs.

%ladder

\begin{figure}
    \begin{subfigure}{0.23\textwidth}
        \subcaptionbox{\label{label1}}
        {\begin{tikzpicture}[baseline, yshift=-50pt]
            \SetDistanceScale{0.65}
            \Vertices[size=.04, opacity=.9, style={shading=ball,blue}]{Data/AFM_triangularvertices_residual.csv}    
            \Edges[RGB]{Data/AFM_triangularedges.csv}    
        \end{tikzpicture}}
        \pgfplotscolorbardrawstandalone[colormap={}{of colormap={reds,},},
        point meta min=0,
        point meta max=1,
        colorbar horizontal,
        colormap access=map,
        colorbar style={
        width=2cm, height=0.1cm, font=\tiny},
        ]
    \end{subfigure}
    ~
    \begin{subfigure}{0.23\textwidth}
        \subcaptionbox{\label{label2}}
        {\begin{tikzpicture}[baseline]
            \begin{axis}[xlabel = $\omega_m/2\pi$ (MHz), ylabel = mode, small, width=4.7cm]
                \addplot+ [ycomb, mark size = 1.5pt, mark=triangle*, dashed] coordinates {(2.36,20)};
                \addplot+ [ycomb, mark size = 1pt] table [x index=1, y index=0] {Data/Spectra_Triangular_lattice_uneven_AFM.dat};
                \addplot+ [ycomb, mark size = 1pt, mark options={green!80!black}, draw=green!80!black] table [x index=2, y index=0] {Data/Spectra_Triangular_lattice_uneven_AFM.dat};
            \end{axis}
            %\begin{axis}[colorbar horizontal, hide axis, width=3cm, height=3cm, xshift=10pt, yshift = 100pt, colorbar style={height=5pt, width=30pt}]
            \begin{axis}[
                hide axis,
                tiny,
                width=2.5cm,
                height=2.5cm,
                xshift=10pt,
                yshift = 40pt,
                colormap/redyellow,
                colorbar horizontal,
                colorbar style={height=3pt, width=23pt, yshift=5pt,xshift=0pt}]
                \addplot+ [scatter, scatter src=explicit, only marks, mark options={draw=none}, mark size = 1.5pt, colormap/redyellow] table [x index=1, y index=2, meta index=3] {Data/Triangular_lattice_uneven_AFM_tweezer_strength.dat};
            \end{axis}
        \end{tikzpicture}}
    \end{subfigure}
    ~
    \begin{subfigure}{0.23\textwidth}
        \subcaptionbox{\label{label1}}
        {\begin{tikzpicture}[baseline, yshift=-50pt]
            \SetDistanceScale{0.65}
            \Vertices[size=.04, opacity=.9, style={shading=ball,blue}]{Data/Triangular_lattice_AFM_19ions_errorvertices.csv}    
            \Edges[RGB]{Data/Triangular_lattice_AFM_19ions_erroredges.csv}    
        \end{tikzpicture}}
        \pgfplotscolorbardrawstandalone[colormap={}{of colormap={errorTL,},},
        point meta min=0,
        point meta max=0.33,
        colorbar horizontal,
        colormap access=map,
        colorbar style={
        width=2cm, height=0.1cm, font=\tiny},
        ]
    \end{subfigure}
    ~
    \begin{subfigure}{0.23\textwidth}
        \subcaptionbox{\label{label1}}
        {\begin{tikzpicture}[baseline, yshift=-50pt]
            \SetDistanceScale{0.65}
            \Vertices[size=.04, opacity=.9, style={shading=ball,blue}]{Data/Triangular_lattice_AFM_19ions_errorvertices_residual.csv}    
            \Edges[RGB]{Data/Triangular_lattice_AFM_19ions_erroredges_residual.csv}    
        \end{tikzpicture}}
        \pgfplotscolorbardrawstandalone[colormap={}{of colormap={errorTLres,},},
        point meta min=0,
        point meta max=0.1,
        colorbar horizontal,
        colormap access=map,
        colorbar style={
        width=2cm, height=0.1cm, font=\tiny, /pgf/number format/fixed, xtick={0,0.05,0.1}},
        ]
    \end{subfigure}

    \caption{\label{fig:triangular_lattice} (a) Resulting frustrated triangular lattice of 19 ions with anti-ferromagnetic couplings. (b) Mode spectra of modified ($\textcolor{green}{\bullet}$) and native ($\scriptscriptstyle\textcolor{red}{\blacksquare}$) phonon modes. The Raman beatnote is indicated by ($\scriptstyle\textcolor{blue}{\blacktriangle}$). Inset: The optical tweezer strength ($\Omega_x^{(i)}/2\pi$ (MHz)) at each ion. Deviation in coupling strength (\(\lvert  J_\text{T,i,j} -\tilde{J}_{i,j} \rvert\)) for (c) nearest neighbor couplings and (d) residual couplings.}
\end{figure}
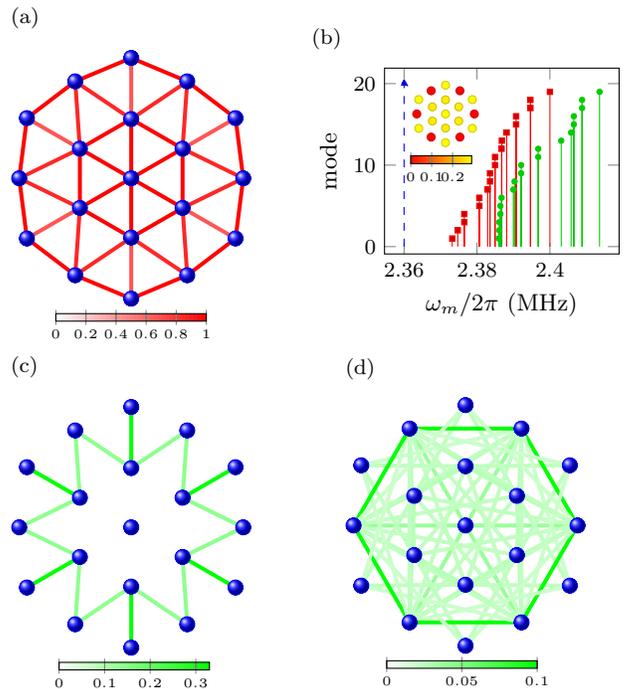

The transverse mode structures are suitable for generating interactions with uniform sign structure. In Fig.~\ref{fig:triangular_lattice}, we illustrate how the combination of these modes and the pinning potential can be used to realize the frustrated triangular lattice with antiferromagnetic nearest-neighbor interactions between 19 ions. 

The relevant experimental parameters of the optimal configuration for both scenarios in 2D are summarized in Table~\ref{tab:summary_2D}.

% Triangular lattice figure

\section{Experimental considerations}\label{sec:exppar}

The dipole interaction due to the optical tweezers can only effectively alter the phonon spectrum of a trapped ion crystal if it can compete with the monopole interaction due to the Paul trap, which is in general dominating. Therefore, it is beneficial to reduce the strength of the Paul trap as much as possible while maintaining the validity of the LD approximation, $\eta_j^{(m)} \ll 1$ for all $j,m$. This sets a lower limit on the strength of the Paul trap. For a single $^{171}$Yb$^+$ ion excited by a pair of Raman beams near the D1 transition at 369~nm with an angle of 90$^{\circ}$ between them and 45$^{\circ}$ with respect to the ion motion, we get $\eta\sim$~0.2 at $\omega_m = 2\pi\cdot$~400~kHz. We note that reducing the angle between the Raman beams further allows us to be deep in the LD regime at even lower phonon frequencies. However this is at the expense of reducing the speed of the quantum simulator in order to eliminate off-resonant coupling. Hence we require optical potentials with local trap frequencies in the 100-400~kHz range to make meaningful changes to the phonon spectrum while maintaining the LD regime.

In order to test the feasibility of changing the local trap frequency on this scale, we estimate the effect of a single tweezer. For $^{171}$Yb$^+$, and an experimentally feasible tweezer power of $P = 1$~W, a waist of $W_0 = 1$~$\upmu$m and wavelength $\lambda = 1070$~nm, we find that the local trap frequencies due to the tweezers are $\sim 2\pi\cdot$~200~kHz~\cite{Grimm:2000,Roy2017}. 

Each optical tweezer will introduce a differential ac-Stark shift between the qubit states of the pinned ions and lead to off-resonant scattering. Both processes can lead to phase shifts \cite{Uys2010} on the state of the system which affect the performance of quantum operations \cite{Erhard2019} and can lead to decoherence. In the case of \(^{171}\text{Yb}^+\), we are concerned about both effects on the qubit states encoded on the hyperfine levels \(\ket{\text{F}=0,m_\text{F}=0}\equiv\ket{0}\) and \(\ket{\text{F}=1,m_\text{F}=0}\equiv\ket{1}\) of the \(^2\text{S}_{1/2}\) ground state.

The photon scattering rate in the center of the Gaussian tweezer beam with waist $W_0$ and peak intensity $I(0) = 2P/\pi W_0^2$ with total light power $P$ is approximately~\cite{Grimm:2000}:
\begin{equation}
    \Gamma_{sc}(\omega) = \frac{3 c^2}{ \hbar\omega_0^3} \left(\frac{\omega}{\omega_0}\right)^3 \left(\frac{\Gamma}{\omega_0 - \omega} + \frac{\Gamma}{\omega_0 + \omega}\right)^2 \frac{P}{W_0^2}.
\end{equation} 

\begin{figure}
    \begin{subfigure}{0.23\textwidth}
        \subcaptionbox{\label{label1}}
        {\begin{tikzpicture}
            \begin{axis}[
                width=4.5cm,
                ymode=log,
                xtick = {250,500,750,1000},
                xlabel={$\lambda\; (\text{nm})$},
                ylabel={$\Gamma/2\pi\; (\text{Hz})$}
                ]
                \addplot [mark=none, very thick] table[x index=0, y index=1, col sep=comma]{Data/photonscatter_vs_wavelength.dat};
            \end{axis}
        \end{tikzpicture}}
    \end{subfigure}
    \begin{subfigure}{0.23\textwidth}
        \subcaptionbox{\label{label2}}
        {\begin{tikzpicture}[baseline]
            \begin{axis}[
                width=4.5cm,
                xlabel = $\Delta_{\text{AC}}/2\pi\;(\text{kHz})$,
                ylabel = $\Omega_p/2\pi\; (\text{kHz})$,
                xtick = {0,6,12}
                ]
                \addplot+[mark=none, very thick] table[x index=0, y index=1, col sep=comma]{Data/difAC_vs_TFtweezer.dat};
            \end{axis}
        \end{tikzpicture}}
    \end{subfigure}
    \caption{\label{fig:scattering_rate}(a) Scattering rate of Yb$^{+1}$, (b) differential Stark shifts and pinning frequencies from a tightly focused tweezer (\(w_0=2\; \upmu\)m) at 1070 nm for a laser power of (a) 1 W and (b) 10 to 1000 mW and linear polarization.}
\end{figure}
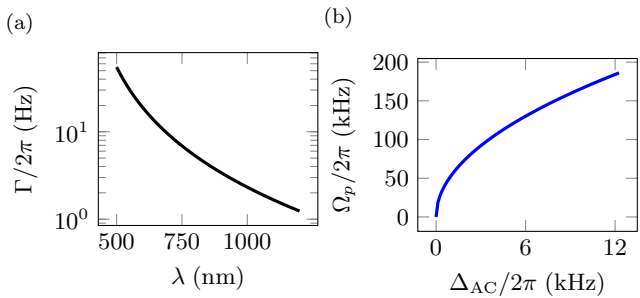

We only consider contributions from the D1 and D2 transitions in $^{171}$Yb$^+$. For $P = 1$~W, $W_0 = 1$~$\upmu$m and wavelength $\lambda = 1070$~nm, the photon scattering rate is $\sim$~2~s$^{-1}$. In Fig.~\ref{fig:scattering_rate}(a), we give an overview of the photon scattering rate at various tweezer wavelengths. Our results indicate that it is possible to modify the phonon spectrum of trapped ions within the LD regime and maintain negligible photon scattering probability on timescales of~$\sim$~100~ms. 

For our qubit states, the differential Stark shift is highly suppressed as compared to the common Stark shift~\cite{Grimm:2000,Lee:2005}. This is because the huge (100 THz) spin-orbit splitting of the P states does not play a role for the hyperfine clock states. The differential Stark shift is almost independent of the tweezer polarization and is dominated by the 12.6~GHz hyperfine splitting that causes a difference in detuning between the two qubit states~\cite{Hirzler:2020,Teoh2021}. For $P = 1$~W, $W_0 = 1 $~$\upmu$m, and $\lambda = 1070$~nm, we obtain a differential Stark shift of $\sim$~$2\pi$~12~kHz (see Fig.~\ref{fig:scattering_rate}(b)). 

Since the tweezer pattern is in general not homogeneous, the differential Stark shift will vary between the ions and will appear as an inhomogeneous additional field in the quantum simulator. The differential Stark shift may be canceled by using pairs of blue and red detuned tweezers such that the combined differential Stark shift becomes zero. Unfortunately, the D1 and D2 transitions for most relevant ions lie in the UV range, making this solution technically demanding. 
%In Fig.~\ref{fig:stark_compensation} we show an example using tweezers operating at 532 and 224 nm. 
Eliminating the \emph{variation} in differential Stark shifts can be done with a single wavelength. This is because the tweezer trap frequency around the $i$-th ion, $I_i=I(\mathbf{r}_i)$ scales as $\Omega^2_i\propto P_i/w_0^4$, whereas the differential Stark shift scales as $\Delta_\text{AC}\propto P_i/w_0^2$. Therefore, we can make the differential Stark shift between the qubit states homogeneous throughout the ion crystal by controlling not only the power $P_i$ of each tweezer but also each waist $w_i$ and assuring that $P_i/w_i^2$ is constant around each ion. While this solution still gives us complete freedom to engineer each local tweezer trap frequency, it comes at the expense of having in general stronger laser power requirements.

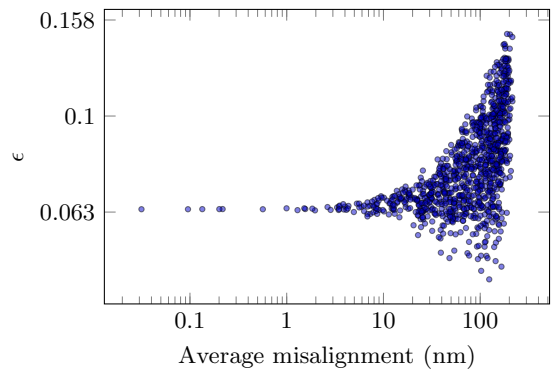
\begin{figure}
    \begin{tikzpicture}
        \begin{loglogaxis}[
        log ticks with fixed point,
        width = 7.5cm,
        height = 5.5cm,
        yticklabels={0.0,0.063,0.1,0.158},
        ylabel=$\epsilon$,
        xlabel=Average misalignment (nm),
        ]
            \addplot +[only marks,
            black,
            opacity=0.5,
            mark size=1pt,
            ] table[x index=0, y index=1, col sep=comma] {Data/dist_error_NN.csv};
        \end{loglogaxis}
    \end{tikzpicture}
    \caption{\label{fig:tweezer_misalign}  Calculated error $\epsilon$ for nearest neighbour interactions and an optimized tweezer pattern that is misaligned. For this we use the tweezer and beatnote frequencies as obtained from our optimizer but allow for misalignments of the tweezers. We have chosen 1000 random tweezer misalignments and plotted the error as a function of the average distance of the tweezer centers to the ion equilibrium positions. }
\end{figure}

Our final source of error is the misalignment of the tweezers from the equilibrium positions of the ions. In this situation, the tweezers start to supply local stress to the ion crystal and for each tweezer setting, new ion equilibrium positions have to be found before the eigenmodes can be obtained. In Fig.~\ref{fig:tweezer_misalign} we show the effect of this type of error on the nearest-neighbour interaction pattern for 12 ions in a linear crystal as also shown in Fig.~\ref{fig:homogeneous_coupling} for perfect alignment. We conclude that our scheme is robust to such misalignments as only  minor deviations from the desired spin-spin interactions occur as long as the alignment can be done to within a resolution of $\sim$~100~nm. Note that for some of the random misalignments an improvement can even be seen, suggesting that using tweezers to apply local forces to the ion crystal may be another useful way of controlling spin-spin interactions.  For these calculations, we approximated the tweezers as harmonic, which is justified as the misalignment $\ll w_0$.

% Misalignment

% Stark shift and pin freq 1070 nm

\section{Discussion}

In this paper we have shown that engineered phonon modes can be used to create a wide variety of interaction ranges and connectivities in trapped-ion quantum simulators. We have shown how optimal tweezer settings can be found using numerical optimization techniques and gave examples of calculated spin-spin interactions in 1D and 2D ion crystals. While we have limited our discussion to Ising-like interactions terms, the optical tweezers can be used to implement spin-spin interactions of the more generic form. This may pave the way for simulating challenging spin models with frustration in triangular, hexagonal and kagome lattices~\cite{Bermudez2011,Bermudez2012,Nath2015} in 2D systems.

The scheme can be straightforwardly combined with other methods developed for engineering spin-spin interactions such as employing multiple beatnote frequencies in the Raman laser to design spin-spin interaction patterns~\cite{Shapira2020, Teoh2020}. Furthermore, the scheme may be extended to let the tweezers apply local stress or strain on the ion crystal modifying the equilibrium positions of the ions. Rapid dynamical control of the tweezers may open up new opportunities in trapped ion quantum information processing~\cite{Olsacher2021,Teoh2021}.

\acknowledgements
We thank Matthias Peschke for fruitful discussions on the optimization method. This work was supported by the Netherlands Organization for Scientific Research (Grant Nos. 680.91.120 and 680.92.18.05, R.G., M.M. and R.X.S.). P.C. and J.D.A.E acknowledge funding from the European Research Council (ERC) under the European Union's Horizon 2020 research and innovation programme (grant agreement No 677061).
A.S.N is supported by the Dutch Research Council (NWO/OCW), as part of the Quantum Software Consortium programme (project number 024.003.037).

\appendix
\section{Gradient coupling matrix}

We can calculate the gradient of the coupling matrix with respect to the Hessian matrix, i.e. \(\bar{\mathbf{A}} = \frac{\partial \mathbf{J}}{\partial \mathbf{A}}\) using adjoint backpropagation methods \cite{Giles2008}. The gradients of each entry $J^{kl}$ with respect to $\mathbf{A}$ is obtained from:

\begin{align}
    \mathbf{\bar{A}}_{kl} = \mathbf{U}_{kl}(\mathbf{\bar{\Lambda}}_{kl} + \frac{\mathbf{F}}{2} \circ (\mathbf{U}_{kl}^\text{T}\mathbf{\bar{U}}_{kl} - \mathbf{\bar{U}}_{kl}^\text{T}\mathbf{U}_{kl}))\mathbf{U}_{kl}^\text{T}
\end{align}

\noindent
where \(\circ\) is the Hadamard product, $\mathbf{U\Lambda U}^\text{T} = \mathbf{A}$, $F^{ij}=(\lambda_j-\lambda_i)^{-1}$ if $i \neq j$ and zero otherwise and the corresponding adjoints of $\mathbf{U}$ and $\mathbf{\Lambda}$ are:

\begin{align*}
    \bar{U}^{ij}_{kl} &= \Theta(j) U^{ij}\;, &(k=i \lor l=i)\\    
    & = 0\;, &(\text{otherwise})\\
    \bar \Lambda^{ii}_{kl} &= \Theta(i)^2U^{ki}U^{li},
\end{align*}

\noindent where \(\Theta(j)=(\mu^2-\lambda_j)^{-1}\). Finally, we write:

\begin{align*}
    \mathbf{\bar{A}} = \left(\text{vec}(\mathbf{\bar A}_{11}), ... \text{vec}(\mathbf{\bar A}_{1N}), ... \text{vec}(\mathbf{\bar A}_{NN}) \right)^\text{T}.
\end{align*}

\bibliographystyle{apsrev4-2}
\bibliography{TweezerIon}

\end{document}